\documentclass[useAMS,usenatbib]{mn2e}

\usepackage[T1]{fontenc}
 \usepackage{pslatex}
\usepackage{amsmath}
\usepackage{amsfonts}
\usepackage{color}
\usepackage{url}
\usepackage{bm}
\usepackage{graphicx}
\usepackage{subfigure}
\usepackage{amssymb}
\usepackage{soul}
\usepackage{booktabs}
\usepackage{array}
\usepackage[utf8]{inputenc}
\inputencoding{latin1}
\inputencoding{utf8}

{\newif\ifnotend
\notendtrue
\def\veclist{ABCDEFGHIJKLMNOPQRSTUVWXYZabcdefghijklmnopqrstuvwxyz.}
\def\top#1#2.{#1}
\def\tail#1#2.{#2.}
\loop\expandafter\xdef\csname v\expandafter\top\veclist\endcsname%
{{\noexpand\bf\expandafter\top\veclist}}
\edef\veclist{\expandafter\tail\veclist}
\if\veclist.\notendfalse\fi\ifnotend\repeat}
\def\e{{\rm e}}

\def\E{{\cal E}}

\mathchardef\mhyphen="2D

\title[Instability of highly magnetised outflows]{Instability induced by recollimation in highly magnetised outflows}
\author[Sobacchi \& Lyubarsky]{E. Sobacchi$^{1,2}$\thanks{E-mail: sobacchi@post.bgu.ac.il} \& Y. E. Lyubarsky$^1$\\
$^1$ Physics Department, Ben-Gurion University, P.O.B. 653, Beer-Sheva 84105, Israel \\
$^2$ Department of Natural Sciences, The Open University of Israel, 1 University Road, P.O.B. 808, Raanana 4353701, Israel
}
\begin{document}

\date{}

\def\p{\partial}
\def\E{\textbf{E}}
\def\B{\textbf{B}}
\def\v{\textbf{v}}
\def\j{\textbf{j}}
\def\s{\textbf{s}}
\def\e{\textbf{e}}

\newcommand{\di}{\mathrm{d}}
\newcommand{\bfx}{\mathbf{x}}
\newcommand{\bfe}{\mathbf{e}}
\newcommand{\vlos}{\mathrm{v}_{\rm los}}
\newcommand{\Tspin}{T_{\rm s}}
\newcommand{\Tb}{T_{\rm b}}
\newcommand{\degree}{\ensuremath{^\circ}}
\newcommand{\Th}{T_{\rm h}}
\newcommand{\Tc}{T_{\rm c}}
\newcommand{\bfr}{\mathbf{r}}
\newcommand{\bfv}{\mathbf{v}}
\newcommand{\bfu}{\mathbf{u}}
\newcommand{\pc}{\,{\rm pc}}
\newcommand{\kpc}{\,{\rm kpc}}
\newcommand{\Myr}{\,{\rm Myr}}
\newcommand{\Gyr}{\,{\rm Gyr}}
\newcommand{\kms}{\,{\rm km\, s^{-1}}}
\newcommand{\de}[2]{\frac{\partial #1}{\partial {#2}}}
\newcommand{\cs}{c_{\rm s}}
\newcommand{\rb}{r_{\rm b}}
\newcommand{\rqu}{r_{\rm q}}
\newcommand{\bfOmega}{\pmb{\Omega}}
\newcommand{\bfOmegap}{\pmb{\Omega}_{\rm p}}
\newcommand{\bfXi}{\boldsymbol{\Xi}}

\maketitle

\begin{abstract}
In the paradigm of magnetic acceleration of relativistic outflows, a crucial point is identifying a viable mechanism to convert the Poynting flux into the kinetic energy of the plasma, and eventually into the observed radiation. Since the plasma is hardly accelerated beyond equipartition, MHD instabilities are often invoked to explain the dissipation of the magnetic energy. Motivated by the fast variability that is shown by the gamma-ray flares of both AGN and PWNe, different authors have proposed the Poynting flux to be dissipated in a region where the flow is converging. Here we perform a linear stability analysis of ultra-relativistic, highly magnetised outflows with such a recollimation nozzle, showing that MHD instabilities are indeed induced by the convergence of the flow. The amplitude of the perturbations increases while recollimation gets stronger, and eventually diverges when the flow is focused to a single point. Hence, depending on the geometry of the outflow, instabilities excited while the flow is converging may play an important role to dissipate the magnetic energy of the plasma.
\end{abstract}



\section{Introduction}
\label{sec:introduction}

Relativistic jets occur ubiquitously among a wide variety of astrophysical sources, including Gamma Ray Bursts (GRBs), Active Galactic Nuclei (AGN), Pulsar Wind Nebulae (PWNe) and microquasars. It is widely accepted that these outflows are powered electromagnetically: the plasma slides on the magnetic field lines anchored to the rotating central object (a black hole or a neutron star) and is gradually accelerated by the magnetic tension (e.g. \citealt{Blandford1976, Lovelace1976, BlandfordZnajek1977}).

In the aforementioned scenario the outflow's energy budget is initially dominated by the electromagnetic fields. Hence, it is crucial to understand how the Poynting flux is converted into the kinetic energy of the plasma, and eventually into the observed radiation. Since gradual acceleration is generally inefficient after equipartition between the magnetic and the kinetic energy is reached (e.g. \citealt{Komissarov2007, Komissarov2009, Lyubarski2009, Lyubarski2010, Lyubarski2011, Tchekhovskoy2008, Tchekhovskoy2009, Tchekhovskoy2010}), MHD instabilities are often invoked to destroy the regular structure of the flow and dissipate the magnetic energy (e.g. \citealt{Lyubarski1992, Eichler1993, Spruit1997, Begelman1998, Giannios2006}).

The impact of the MHD instabilities depends on the causality condition in the lateral direction. In this paper we focus on the case of highly magnetised outflows, namely we assume the magnetisation to be $\sigma\gg 1$ (the Lorentz factor corresponding to the fast magnetosonic velocity is then $\sqrt{\sigma}$). If $\theta\gamma\gtrsim \sqrt{\sigma}$, the flow is causally disconnected in the lateral direction and global instabilities cannot develop \citep{Granot2011}. However, the plasma can pass through a weak shock whose only role is to make the downstream flow causally connected (for a more extended discussion see \citealt{Lyubarsky2012}). Hence, one can restrict the stability analysis to the causally connected regime $\theta\gamma\lesssim \sqrt{\sigma}$.

If $\theta\gamma\lesssim 1$ (strong causal connection), the signal crossing time is shorter than the expansion time and the flow structure at any distance from the source is relaxed to an appropriate cylindrical equilibrium configuration \citep{Lyubarski2009}. The residual of the hoop stress and the electric force is balanced by the poloidal magnetic field and the ratio of the toroidal to poloidal components of the magnetic field is $B_\phi/B_{\rm p}\sim\gamma$, namely the two components are of the same order in the proper frame of the plasma.

In the context of cylindrical jets, the most dangerous kink ($m=1$) modes have been investigated by a number of authors, both analytically (e.g. \citealt{IstominPariev1996, Lyubarski1999, Appl2000}) and numerically (e.g. \citealt{Nakamura2007, Mignone2010, Mizuno2012}). It turns out that the kink modes are stabilised if the strength of the poloidal magnetic field is nearly independent of the cylindrical radius $r$ (e.g. \citealt{IstominPariev1996, Lyubarski1999, Mizuno2012, Sobacchi2017}), which might be the case in a realistic scenario \citep{Narayan2009}.

If $1\lesssim\theta\gamma\lesssim \sqrt{\sigma}$ (weak causal connection), the flow expands significantly within a signal crossing time and thus it is not in transverse equilibrium. However, the flow can still be adjusted to the external pressure without passing through a shock. One can show that the toroidal component of the magnetic field is much larger, in the proper frame, than the poloidal one \citep{Lyubarski2009}. Therefore one can consider the flow as composed of coaxial magnetic coils. It is important to realise that the confining pressure of the external medium can induce recollimation nozzles along the flow \citep{Lyubarski2009, GlobusLevinson2016}.

Focusing on the case of PWNe, \citet{Lyubarsky2012} suggested how the relevant instability should work. Consider a region where the magnetic field can be effectively approximated as toroidal, so that the flow can be conceived as composed of coaxial magnetic coils, and take a small radial displacement of two coils. As the flow is converging the relative displacement increases, eventually becoming comparable with the radius of the coil itself. Hence, the regular structure of the field lines is destroyed and the recollimation nozzle provides a viable location to dissipate the magnetic energy.

More recently, numerical simulations (\citealt{BrombergTchekhovskoy2016}; see also \citealt{Barniol2017}) have investigated a setup that might resemble the physical conditions of GRBs, namely a jet launched by a magnetar-like engine and collimated by the pressure of the stellar environment. Interestingly, it was found that (i) the jet generally experiences a rapid expansion that forces $B_\phi/B_{\rm p}\gg\gamma$, and is then recollimated by the external pressure; (ii) the instability indeed develops close to the nozzle. Due to the robustness of such a feature, these authors suggested the instabilities excited while the flow is converging to be the mechanism that powers the GRB prompt emission.

The case of dissipation occurring in the vicinity of a recollimation nozzle is supported by the fact that the gamma-ray emission is often variable on extremely short time scales ($\lesssim\text{day}$ for both typical AGN and PWNe flares, and even down to few minutes in the former case), which poses serious constraints on the size of the emitting region. For this reason, different authors have proposed that the observed flares are produced in a region where the flow is converging, so that its transverse size is reduced; see for example \citet{BrombergLevinson2009, GlobusLevinson2016} in the context of AGN and \citet{Lyubarsky2012} in the context of PWNe, and references therein.

Here we present a linear stability analysis of converging outflows far from a local cylindrical equilibrium, aiming to identify the fundamental physical parameters that regulate the growth of the instability. We find that the growth of the unstable modes is simply determined by the geometry of the outflow, with narrower recollimation nozzles being more unstable, while it is independent of other features such as the Lorentz factor of the flow.

The paper is organised as follows. In Section \ref{sec:equations} we present the fundamental equations governing Poynting-dominated outflows and we find a stationary solution for their structure. In Section \ref{sec:evolution} we study the linear evolution of perturbations propagating along the flow. Finally, in Section \ref{sec:conclusions} we summarise our conclusions.

\section{Fundamental equations}
\label{sec:equations}

The fundamental equations governing magnetised outflows are the Maxwell's equations:
\begin{align}
\label{eq:maxwell_1}
\nabla\times\E & = -\frac{\p\B}{\p t} & \nabla\cdot\B & =0 \\
\label{eq:maxwell_2}
\nabla\cdot\E & = \rho& \nabla\times\B & = \j + \frac{\p\E}{\p t}\;.
\end{align}
We are using Heaviside units and the speed of light is $c=1$.

These equations should combined with the continuity and the Euler's fluid equations. However, if the outflow is weakly causally connected in the lateral direction, the inertia of the plasma may be neglected (e.g. \citealt{Lyubarsky2012}) and Eq. \eqref{eq:maxwell_1}-\eqref{eq:maxwell_2} are simply coupled with the force-free limit of the Euler's fluid equation,
\begin{equation}
\label{eq:euler_forcefree}
\rho\E = -\j\times\B\;.
\end{equation}
If $\rho$ is not vanishing, Eq. \eqref{eq:euler_forcefree} is equivalent to a set of three scalar equations, namely 
\setcounter{equation}{2}
\begin{subequations}
\begin{align}
\label{eq:euler_B}
\E\cdot\B&=0\\
\label{eq:euler_j}
\E\cdot\j&=0\\
\label{eq:euler_sq}
\rho^2\E^2&=\j^2\B^2-\left(\j\cdot\B\right)^2\;.
\end{align}
\end{subequations}
These are obtained from Eq. \eqref{eq:euler_forcefree} respectively (i) projecting along $\B$; (ii) projecting along $\j$; (iii) taking the square of both sides. In particular, note that the ideal MHD condition $\E\cdot\B=0$ is already encoded in Eq. \eqref{eq:euler_forcefree}.

\subsection{Unperturbed solution}

We first look for a stationary solution of Eq. \eqref{eq:maxwell_1}-\eqref{eq:euler_forcefree}. We consider a strongly expanded, non-cylindrical outflow far from a local cylindrical equilibrium (in the sense discussed in the introduction), and we therefore neglect any poloidal component of the magnetic field. For the sake of simplicity, we take the pressure of the confining medium to be constant.

A stationary solution for such a configuration was already presented in \citet{Lyubarsky2012}, which we refer to for a detailed derivation. The outer radius $R$ of the flow evolves as
\begin{equation}
\label{eq:jet_radius}
R\equiv R_{\rm max}\left|\cos\left(\frac{z}{L}\right)\right|\;,
\end{equation}
with $R_{\rm max}\ll L$. The origin $z=0$ corresponds to the maximum expansion $R=R_{\rm max}$, while the flow is focused to $R=0$ at $z=\pm \pi L/2$, so that the oscillation is occurring on a scale $\pi L$ in the $z$ direction. Note that the field lines are converging (diverging) when $\tau\equiv\tan\left(z/L\right)$ is positive (negative).

The solution for the electromagnetic fields and for the charge and current densities can be parametrised as
\begin{align}
\label{eq:E0}
\E_0 & =  \alpha\frac{r}{R^2}\left[\e_{\rm r} + \tau\frac{r}{L}\e_{\rm z}\right]\\
\label{eq:B0}
\B_0 & =  \alpha\frac{r}{R^2}\left[1 + \frac{1}{6}\left(1+3\tau^2\right)\frac{r^2}{L^2}\right]\e_\phi\\
\label{eq:rho0}
\rho_0 & =  \frac{\alpha}{R^2}\left[2 + \left(1+3\tau^2\right)\frac{r^2}{L^2}\right]\\
\label{eq:j0}
\j_0 & =  \frac{\alpha}{R^2}\left[-2\tau\frac{r}{L}\e_{\rm r} + \left[2+\frac{2}{3}\left(1+3\tau^2\right)\frac{r^2}{L^2}\right]\e_{\rm z}\right]\;,
\end{align}
where $\alpha$ is a scaling constant and $\tau\equiv\tan\left(z/L\right)$. The Lorentz factor corresponding to the drift velocity $\v=\E_0\times\B_0/B_0^2$ is given by
\begin{equation}
\label{eq:drift}
\gamma^2=\frac{B_0^2}{B_0^2-E_0^2}\approx 3\frac{L^2}{r^2}\;,
\end{equation}
and $R\ll L$ is therefore required for the flow to be ultra-relativistic, namely the outer radius needs to be much shorter than the oscillation scale along $z$.

It is simple to verify a posteriori that the fields presented above are indeed a solution of Eq. \eqref{eq:maxwell_1}-\eqref{eq:euler_forcefree} accurate up to terms of the order of $R^2/L^2\sim 1/\gamma^2\ll 1$, and that the electromagnetic pressure at $r=~R$ is constant to the same order. Finally, note that this solution is not valid when $R\to 0$, since at some point the poloidal component of the magnetic field and the thermal pressure of the plasma necessarily become important. In the following we assume this to happen when the outer radius is $R\ll R_{\rm max}$.

\section{Linear evolution of the perturbations}
\label{sec:evolution}

In the following we study the evolution of perturbations that depend on the time and on the azimuthal angle through $\exp\left(-i\omega t + im\phi\right)$, and we focus on non-axisymmetric perturbations. The axisymmetric (i.e. $m=0$) case is analysed in Appendix \ref{sec:appendixA}, where we show these perturbations to be stable. In order to keep the problem analytically treatable, we work in the limit $\omega R\ll\gamma$.

The linearised Faraday's induction equation relates the perturbed magnetic field $\B_1$ to the perturbed electric field $\E_1$:
\begin{equation}
\label{eq:B1}
\B_1=-\frac{i}{\omega}\nabla\times\E_1\;.
\end{equation}
Note that the above relation automatically gives $\nabla\cdot\B_1=0$. From Eq. \eqref{eq:maxwell_2} it is then possible to calculate the perturbed charge and current densities:
\begin{align}
\label{eq:rho1}
\rho_1 & =  \nabla\cdot\E_1\\
\label{eq:j1}
\j_1 & = \nabla\times\B_1 + i\omega \E_1\;.
\end{align}
Note that all the perturbed quantities can be calculated from Eq. \eqref{eq:B1}-\eqref{eq:j1} once $\E_1$ is specified. These equations need to be complemented with the linearised versions of Eq. \eqref{eq:euler_B}-\eqref{eq:euler_sq}, which are
\begin{subequations}
\label{eq:euler_lin}
\begin{align}
\label{eq:EB_lin}
\E_0\cdot\B_1 + \B_0\cdot\E_1 & =0\\
\label{eq:Ej_lin}
\E_0\cdot\j_1 + \j_0\cdot\E_1 & =0\\
\label{eq:eulersq_lin}
\rho_0^2\;\E_0\cdot\E_1-\j_0^2\;\B_0\cdot\B_1 & = \B_0^2\;\j_0\cdot\j_1-\E_0^2\;\rho_0\rho_1\;.
\end{align}
\end{subequations}
While deriving the last equation, we have used the fact that $\left(\j\cdot\B\right)^2$ is a second order quantity since $\j_0\cdot\B_0=0$.

Finally, we are considering the outer radius, $r=R$, as a rigid wall.\footnote{This assumption is equivalent to take the limit where the sound speed of the confining medium is negligibly small with respect to $c$. If a hot medium is allowed, additional modes are excited by the velocity shear at the contact surface between the magnetised plasma and the confining gas. See \citet{SobacchiLyubarsky2018} for a more extended discussion.} The required conditions at the outer boundary are 
\begin{equation}
\label{eq:boundary}
\E_1\times{\bf n}=0 \qquad \B_1\cdot{\bf n}=0\;,
\end{equation}
where
\begin{equation}
{\bf n} = \e_{\rm r} + \tau\frac{R}{L}\e_{\rm z}\;
\end{equation}
is the vector normal to the surface $r=R$. Writing Eq. \eqref{eq:boundary} in components we get
\begin{equation}
\label{eq:outer_boundary}
E_{1\phi}=0 \qquad E_{\rm 1z} - \tau\frac{R}{L}E_{\rm 1r} = 0 \qquad B_{\rm 1r} + \tau\frac{R}{L}B_{\rm 1z}=0\;,
\end{equation}
where all the fields are evaluated at $r=R$. Since ${\bf n}$ is directed along $\E_0$, looking at Eq. \eqref{eq:EB_lin} it is simple to realise that the first and the third equations of \eqref{eq:outer_boundary} are equivalent.

\subsection{Expansion in powers of $\bm{1/\gamma}$}

Since we are considering relativistic flows with $\gamma\sim L/R\gg 1$, we expect perturbations to move at approximately the speed of light. Hence, it is useful to parametrise the perturbations as
\begin{equation}
\label{eq:pert}
f\left(r,z\right)\exp\left[ i\omega \left( z - t \right) + im\phi \right]\;,
\end{equation}
where $\omega$ is a real number and the growth/suppression of the perturbation is encoded in the function $f\left(r,z\right)$. In the limit $L/R\to\infty$, which corresponds to a cylindrical jet with a drift velocity of $c$, we expect $f$ to become independent of $z$.

In the following we are neglecting quantities of order higher than $1/\gamma^2\sim R^2/L^2$, which is the same procedure adopted to find the unperturbed solution. Since the typical length over which the unperturbed fields vary along $z$ is $L$, adopting the parametrisation of Eq. \eqref{eq:pert} we expect $R\partial f/\partial z \sim fR/L \sim f/\gamma$.

\subsubsection{Fundamental scalings}
\label{sec:scalings}

Our next goal is understanding how the different fields and their combinations scale with the small parameter $1/\gamma$. In the following we leave the dependence of the fields on the exponential factor of Eq. \eqref{eq:pert} implicit. Eq. \eqref{eq:EB_lin} can be written as
\begin{equation}
\label{eq:Br}
B_{\rm 1r}+E_{1\phi}=-\frac{\tau r}{L}B_{\rm 1z}-\frac{1}{6}\left(1+3\tau^2\right)\frac{r^2}{L^2}E_{1\phi}\;,
\end{equation}
which implies that $B_{\rm 1r}+E_{1\phi}$ is small. The $r$ component of Eq. \eqref{eq:B1} gives
\begin{equation}
\label{eq:Ez}
E_{\rm 1z}=-i\frac{r}{m}\frac{\partial E_{1\phi}}{\partial z}+\frac{\omega r}{m}\left(B_{\rm 1r}+E_{1\phi}\right)\;.
\end{equation}
Note that the second term on the right hand side is a small correction due to Eq. \eqref{eq:Br}, and $E_{\rm 1z}$ is therefore of the order of $1/\gamma$. From the $z$ component of Eq. \eqref{eq:B1} we find
\begin{equation}
\label{eq:diff1}
\frac{\partial}{\partial r}\left(rE_{1\phi}\right)=imE_{\rm 1r}+i\omega rB_{\rm 1z}\;.
\end{equation}
Using Eq. \eqref{eq:Ez} to express $E_{\rm 1z}$ and then Eq. \eqref{eq:diff1}, the $\phi$ component of Eq. \eqref{eq:B1} can be written as
\begin{equation}
\label{eq:diff3}
B_{1\phi}-E_{\rm 1r}=i\frac{r}{m}\frac{\partial B_{\rm 1z}}{\partial z}+\frac{i}{m}\frac{\partial}{\partial r}\left[r\left(B_{\rm 1r}+E_{1\phi}\right)\right]\;.
\end{equation}
Therefore $B_{1\phi}-E_{\rm 1r}$ is small.

Neglecting terms of the order of $E_{\rm 1z}/\gamma^2$, Eq. \eqref{eq:Ej_lin} gives
\begin{equation}
\label{eq:jr}
rj_{\rm 1r} = -\frac{\tau r}{L}rj_{\rm 1z} + 2\frac{\tau r}{L}E_{\rm 1r} -2E_{\rm 1z}\;.
\end{equation}
Hence, also $j_{\rm 1r}$ is of the order of $1/\gamma$. The $r$ component of Eq. \eqref{eq:j1} gives
\begin{equation}
\label{eq:Bz}
B_{\rm 1z}=\frac{1}{m}\left[-ir\frac{\partial B_{1\phi}}{\partial z} - irj_{\rm 1r} + \omega r\left(B_{1\phi}-E_{\rm 1r}\right)\right]\;.
\end{equation}
Combining Eq. \eqref{eq:jr} and \eqref{eq:Bz}, one can realise that $B_{\rm 1z}$ is of the order of $1/\gamma$. Hence, from Eq. \eqref{eq:Br} and \eqref{eq:diff3} we see that both $B_{\rm 1r}+E_{1\phi}$ and $B_{1\phi}-E_{\rm 1r}$ are of the order of $1/\gamma^2$. The $\phi$ component of Eq. \eqref{eq:j1} can be simply regarded as a definition of $j_{1\phi}$, which does not appear elsewhere. Using Eq. \eqref{eq:rho1}, the $z$ component of Eq. \eqref{eq:j1} can be finally presented as
\begin{equation}
\label{eq:diff4}
r\left(j_{\rm 1z}-\rho_1\right)=\frac{\partial}{\partial r}\left[r\left(B_{1\phi}-E_{\rm 1r}\right)\right] -im\left(B_{\rm 1r}+E_{1\phi}\right) -r\frac{\partial E_{\rm 1z}}{\partial z}\;,
\end{equation}
which shows that also $j_{\rm 1z}-\rho_1$ is of the order of $1/\gamma^2$.

We can eventually summarise how the different fields and their combinations scale with the typical Lorentz factor of the flow:
\begin{align}
\label{eq:scaling1}
E_{\rm 1r} \sim E_{1\phi}\sim B_{\rm 1r}\sim B_{1\phi}\sim r\rho_1\sim rj_{\rm 1z} & \sim 1\\
\label{eq:scaling2}
E_{\rm 1z} \sim B_{\rm 1z}\sim rj_{\rm 1r} & \sim 1/\gamma\\
\label{eq:scaling3}
B_{\rm 1r} + E_{1\phi} \sim B_{1\phi} - E_{\rm 1r} \sim r\left(j_{\rm 1z}-\rho_1\right) & \sim 1/\gamma^2\;.
\end{align}
Note that in the ultra-relativistic limit the perturbed fields almost entirely lie in the $r-\phi$ plane. Since the perturbed magnetic field does not have any $z$ component, one may think to these perturbations as small radial displacements of otherwise coaxial magnetic coils.

\subsubsection{The missing equations}

We have still not used Eq. \eqref{eq:rho1} and Eq. \eqref{eq:eulersq_lin}. From Eq. \eqref{eq:rho1} we get
\begin{equation}
\label{eq:diff2}
\frac{\partial}{\partial r}\left(rE_{\rm 1r}\right)=r\rho_1-imE_{1\phi} -i\omega rE_{\rm 1z} -r\frac{\partial E_{\rm 1z}}{\partial z}\;.
\end{equation}
Neglecting terms of order higher than $1/\gamma^2$, Eq. \eqref{eq:eulersq_lin} gives
\begin{align}
& 2\left(E_{\rm 1r}-B_{1\phi}\right) + 2\frac{\tau r}{L}E_{\rm 1z} + \frac{1}{3}\left(1-3\tau^2\right)\frac{r^2}{L^2}E_{\rm 1r}=\nonumber\\
& = r\left(j_{\rm 1z}-\rho_1\right) -\frac{\tau r}{L}rj_{\rm 1r} + \frac{1}{6}\left(1-3\tau^2\right)\frac{r^2}{L^2}r\rho_1\;,
\end{align}
which, after substituting $rj_{\rm 1r}$ from Eq. \eqref{eq:jr}, simplifies to
\begin{equation}
\label{eq:rho}
r\left(j_{\rm 1z}-\rho_1\right) = 2\left(E_{\rm 1r}-B_{1\phi}\right) + \frac{1}{6}\left(1+3\tau^2\right)\frac{r^2}{L^2}\left(2E_{\rm 1r}-r\rho_1\right)
\end{equation}
Since all the terms in Eq. \eqref{eq:rho} are of the order of $1/\gamma^2$, expanding the solution to this order is necessary to avoid the problem to become degenerate.

\subsubsection{Further manipulations}
\label{sec:manipulations}

It is now possible to rewrite the equations obtained in the last two sections keeping only the leading terms. Terms of order unity in Eq. \eqref{eq:diff1} and \eqref{eq:diff2} give respectively
\begin{align}
\frac{\partial}{\partial r}\left(rE_{1\phi}\right) & =imE_{\rm 1r}\label{eq:first}\\
\frac{\partial}{\partial r}\left(rE_{\rm 1r}\right) & =r\rho_1-imE_{1\phi}\;.
\end{align}
Terms of the order of $1/\gamma$ in Eq. \eqref{eq:Ez}, \eqref{eq:jr} and \eqref{eq:Bz} give
\begin{align}
E_{\rm 1z} & =-i\frac{r}{m}\frac{\partial E_{1\phi}}{\partial z}\\
rj_{\rm 1r} & = -\frac{\tau r}{L}r\rho_1 + 2\frac{\tau r}{L}E_{\rm 1r} -2E_{\rm 1z} \label{eq:j1r_1}\\
B_{\rm 1z} & = -\frac{i}{m}\left[r\frac{\partial E_{\rm 1r}}{\partial z} + rj_{\rm 1r}\right]\;. \label{eq:j1r_2}
\end{align}
Finally, terms of the order of $1/\gamma^2$ in Eq. \eqref{eq:Br}, \eqref{eq:diff3}, \eqref{eq:diff4} and \eqref{eq:rho} give respectively
\begin{align}
B_{\rm 1r}+E_{1\phi} & =-\frac{\tau r}{L}B_{\rm 1z}-\frac{1}{6}\left(1+3\tau^2\right)\frac{r^2}{L^2}E_{1\phi}\\
B_{1\phi}-E_{\rm 1r} & =i\frac{r}{m}\frac{\partial B_{\rm 1z}}{\partial z}+\frac{i}{m}\frac{\partial}{\partial r}\left[r\left(B_{\rm 1r}+E_{1\phi}\right)\right]\label{eq:thirdlast}\\
r\left(j_{\rm 1z}-\rho_1\right) & =\frac{\partial}{\partial r}\left[r\left(B_{1\phi}-E_{\rm 1r}\right)\right] -im\left(B_{\rm 1r}+E_{1\phi}\right) -r\frac{\partial E_{\rm 1z}}{\partial z}\label{eq:secondlast}\\
r\left(j_{\rm 1z}-\rho_1\right) & = 2\left(E_{\rm 1r}-B_{1\phi}\right) + \frac{1}{6}\left(1+3\tau^2\right)\frac{r^2}{L^2}\left(2E_{\rm 1r}-r\rho_1\right)\label{eq:last}
\end{align}
Note that Eq. \eqref{eq:first}-\eqref{eq:last} is a set of nine equations for the nine unknown functions $E_{\rm 1r}$, $E_{1\phi}$, $E_{\rm 1z}$, $B_{\rm 1r}+E_{1\phi}$, $B_{1\phi}-E_{\rm 1r}$, $B_{\rm 1z}$, $j_{\rm 1r}$, $\rho_1$, $j_{\rm 1z}-\rho_1$.

It is convenient to introduce the new variables $\tau\equiv\tan\left(z/L\right)$ and $s\equiv -\log\left(r/R\right)$, where $R\equiv R_{\rm max}\left|\cos\left(z/L\right)\right|$ is the outer radius of the flow. This gives the following chain rule for the partial derivatives:
\begin{align}
\label{eq:chain1}
\frac{\partial}{\partial r} & =\frac{\partial\tau}{\partial r}\frac{\partial}{\partial\tau} + \frac{\partial s}{\partial r}\frac{\partial}{\partial s}= -\frac{1}{r}\frac{\partial}{\partial s}\\
\label{eq:chain2}
\frac{\partial}{\partial z} & =\frac{\partial\tau}{\partial z}\frac{\partial}{\partial\tau} + \frac{\partial s}{\partial z}\frac{\partial}{\partial s}=\frac{1}{L}\left(1+\tau^2\right)\frac{\partial}{\partial\tau} - \frac{1}{L} \tau \frac{\partial}{\partial s} \;.
\end{align}
Note that in the new coordinates the axis $r=0$ corresponds to the limit $s\to\infty$, while the outer radius of the flow $r=R$ corresponds to $s=0$.

We may now present Eq. \eqref{eq:first}-\eqref{eq:last} using the new definitions: $E_{\rm 1r}\equiv X$, $E_{1\phi}\equiv imY$, $r\rho_1\equiv Q$, $E_{\rm 1z}\equiv \left(\tau r/L\right)E$, $B_{\rm 1z}\equiv im\left(\tau r/L\right)B$, $B_{\rm 1r}+E_{1\phi}\equiv im\left(\tau r/L\right)^2 U$, $B_{1\phi}-E_{\rm 1r}\equiv \left(\tau r/L\right)^2 V$. Substituting Eq. \eqref{eq:j1r_1} into Eq. \eqref{eq:j1r_2} to eliminate $rj_{\rm 1r}$, from Eq. \eqref{eq:first}-\eqref{eq:thirdlast} we get
\begin{align}
X & = Y - \frac{\partial Y}{\partial s} \label{eq:1}\\
Q & = X - \frac{\partial X}{\partial s} -m^2 Y \label{eq:2}\\
E & = \left(1+\frac{1}{\tau^2}\right)\tau\frac{\partial Y}{\partial \tau} - \frac{\partial Y}{\partial s} \label{eq:4}\\
m^2 B & =-\left(1+\frac{1}{\tau^2}\right)\tau\frac{\partial X}{\partial \tau} +\frac{\partial X}{\partial s} + Q -2 X +2E \label{eq:6}\\
U & = -B - \frac{1}{6}\left(3+\frac{1}{\tau^2}\right)Y \label{eq:3}\\
V & =-\left(1+\frac{1}{\tau^2}\right)\left[B+\tau\frac{\partial B}{\partial \tau}\right] +\frac{\partial B}{\partial s} -3U + \frac{\partial U}{\partial s} \label{eq:5}\;.
\end{align}
All the functions can be expressed in terms of $Y$ after deriving (i) $X$ from Eq. \eqref{eq:1}; (ii) $Q$ from Eq. \eqref{eq:2}; (iii) $E$ from Eq. \eqref{eq:4}; (iv) $B$ from Eq. \eqref{eq:6}; (v) $U$ from Eq. \eqref{eq:3}; (vi) $V$ from Eq. \eqref{eq:5}. Finally, combining Eq. \eqref{eq:secondlast} and \eqref{eq:last} to eliminate $j_{\rm 1z}-\rho_1$, we find
\begin{align}
\frac{\partial E}{\partial s} -\left(1+\frac{1}{\tau^2}\right) & \left[E+\tau\frac{\partial E}{\partial \tau}\right] -\frac{\partial V}{\partial s} = \nonumber\\
& = \frac{1}{6}\left(3 +\frac{1}{\tau^2}\right)\left(2X-Q\right) -5V -m^2U\;.\label{eq:7}
\end{align}
Substituting all the functions \eqref{eq:1}-\eqref{eq:5} into Eq. \eqref{eq:7}, one would eventually get a partial differential equation for $E_{1\phi}\equiv imY$.

\subsection{The limit $\bm{R\ll R_{\bf max}}$}

In the limit $R\ll R_{\rm max}$, namely close to the recollimation nozzle where the unperturbed flow is approximately conical, the set of equations \eqref{eq:1}-\eqref{eq:7} is significantly simplified due to the fact that $\left|\tau\right|\gg 1$. Defining a new variable $u\equiv\log\left|\tau/\tau_0\right|$, where $\tau_0$ is an arbitrary constant, and neglecting small corrections of the order of $1/\tau^2$, Eq. \eqref{eq:1}-\eqref{eq:7} become
\begin{align}
X & = Y -\frac{\partial Y}{\partial s} \label{eq:first1}\\
Q & = X -\frac{\partial X}{\partial s} -m^2 Y \\
E & =\frac{\partial Y}{\partial u} -\frac{\partial Y}{\partial s} \\
m^2 B & =-\frac{\partial X}{\partial u} +\frac{\partial X}{\partial s} +Q -2X +2E \\
U & = -B -\frac{1}{2}Y \\
V & = -B -\frac{\partial B}{\partial u} +\frac{\partial B}{\partial s} -3U + \frac{\partial U}{\partial s} \\
\frac{\partial E}{\partial s} & -E -\frac{\partial E}{\partial u} -\frac{\partial V}{\partial s} = X- \frac{1}{2}Q -5V -m^2U \label{eq:last1}\;.
\end{align}

Eq. \eqref{eq:first1}-\eqref{eq:last1} should be regarded as an initial value problem, where the amplitude of the perturbed fields is specified on a certain plane $\tau=\tau_0$. Hence, the problem is better solved through the Laplace transform
\begin{equation}
\label{eq:laplace}
\tilde{Y}\left(s,p\right)\equiv\int_0^\infty\text{d}u\operatorname{e}^{-pu}Y\left(s,u\right)\;,
\end{equation}
where $\operatorname{Re}\left(p\right)$ is larger than the fastest growing exponential term contained in $Y$. Analogous definitions hold for the other functions appearing in our equations. Eq. \eqref{eq:first1}-\eqref{eq:last1} become
\begin{align}
\tilde{X} & = \tilde{Y} -\frac{\partial \tilde{Y}}{\partial s} \label{eq:first2}\\
\tilde{Q} & = \tilde{X} -\frac{\partial \tilde{X}}{\partial s} -m^2 \tilde{Y} \\
\tilde{E} & =p\tilde{Y} -\frac{\partial \tilde{Y}}{\partial s} -Y_0\\
m^2 \tilde{B} & =-\left(p+2\right)\tilde{X} +\frac{\partial \tilde{X}}{\partial s} +\tilde{Q} +2\tilde{E} +X_0 \\
\tilde{U} & = -\tilde{B} -\frac{1}{2}\tilde{Y} \\
\tilde{V} & = -(p+1)\tilde{B} +\frac{\partial \tilde{B}}{\partial s} -3\tilde{U} + \frac{\partial \tilde{U}}{\partial s} +B_0 \\
\frac{\partial \tilde{E}}{\partial s} & -\left(p+1\right)\tilde{E} -\frac{\partial \tilde{V}}{\partial s} +E_0 = \tilde{X}- \frac{1}{2}\tilde{Q} -5\tilde{V} -m^2\tilde{U} \label{eq:last2}\;,
\end{align}
where $Y_0\left(s\right)\equiv Y\left(s,0\right)$ is the initial $Y$ field (i.e. it is evaluated on the plane $\tau=\tau_0$), and analogous expressions hold for the other fields.

Using the procedure outlined at the end of Section \ref{sec:manipulations}, we can now rearrange Eq. \eqref{eq:first2}-\eqref{eq:last2} in order to find a single differential equation for $\tilde{Y}$. We finally get
\begin{equation}
\label{eq:final2}
\frac{\partial^2 \tilde{Y}}{\partial s^2} -4\frac{\partial \tilde{Y}}{\partial s} -\left(m^2+5\right)\tilde{Y} = \frac{C_0\left(p,s,m\right)}{p^2-3p+2}\;,
\end{equation}
where the function $C_0$ depends on the initial fields. One can provide two independent modes on the plane $\tau=\tau_0$; the most convenient choice is using $Y_0$ and $Z_0$, which is defined through $Z\equiv\partial Y/\partial u$. This gives
\begin{equation}
C_0\equiv Z_0''-4Z_0' -\left(m^2+5\right)Z_0 +\left(p-3\right)\left[Y_0''-4Y_0' -\left(m^2+5\right)Y_0\right]\;,
\end{equation}
where the prime denotes the derivative with respect to $s$. Note that using $Z_0$ in the definition of $C_0$ is equivalent, for example, to use $E_0=Z_0-Y_0'$.

\subsubsection{Boundary conditions}

Eq. \eqref{eq:final2} needs to be combined with the appropriate boundary conditions. The first and the third equations of \eqref{eq:outer_boundary} are satisfied if
\begin{equation}
\label{eq:bc1}
E_{1\phi}=0\;,
\end{equation} 
where the field is evaluated in $s=0$. Evaluating Eq. \eqref{eq:Br}, \eqref{eq:Ez} and \eqref{eq:diff1} in $r=R$, it is simple to realise that the second equation of \eqref{eq:outer_boundary} is equivalent to
\begin{equation}
\frac{\partial E_{1\phi}}{\partial z} -\frac{\tau r}{L} \frac{\partial E_{1\phi}}{\partial r}=0\;.
\end{equation}
Using Eq. \eqref{eq:chain1}-\eqref{eq:chain2}, this condition reduces to
\begin{equation}
\label{eq:bc2}
\frac{\partial E_{1\phi}}{\partial\tau}=0\;,
\end{equation}
where the field is evaluated in $s=0$. Hence, our boundary condition is simply
\begin{equation}
\label{eq:outer_boundary1}
Y\left(0,u\right)=0\;,
\end{equation}
which guarantees both Eq. \eqref{eq:bc1} and Eq. \eqref{eq:bc2} to be satisfied. Moreover, the regularity of the solution on the axis requires
\begin{equation}
\label{eq:outer_boundary2}
\lim_{s\to+\infty}Y\left(s,u\right)=0\;.
\end{equation}
Using Eq. \eqref{eq:laplace}, it is simple to realise that Eq. \eqref{eq:outer_boundary1}-\eqref{eq:outer_boundary2} are equivalent to
\begin{equation}
\label{eq:BC_final}
\tilde{Y}\left(0,p\right) = 0 \qquad\qquad \lim_{s\to +\infty}\tilde{Y}\left(s,p\right)=0\;,
\end{equation}
which are our final boundary conditions for $\tilde{Y}$.

\subsubsection{Growth of the perturbations}

We finally need to solve Eq. \eqref{eq:final2} with the boundary conditions \eqref{eq:BC_final}. The solution is
\begin{equation}
\label{eq:sol_laplace}
\tilde{Y}\left(s,p\right)=\frac{f\left(s,p\right)}{p^2-3p+2}\;,
\end{equation}
where
\begin{equation}
f\left(s,p\right)\equiv Z_0 + \left(p-3\right)Y_0\;.
\end{equation}
Note that, since both $Y_0$ and $Z_0$ needs to be consistent with Eq. \eqref{eq:outer_boundary1}-\eqref{eq:outer_boundary2}, $\tilde{Y}$ is automatically satisfying our boundary conditions. We can now evaluate $Y\left(s,u\right)$ performing the inverse transformation of Eq. \eqref{eq:sol_laplace}. We get
\begin{equation}
\label{eq:sol_laplace2}
Y\left(s,u\right)=\frac{1}{2\pi i}\int_{\beta-i\infty}^{\beta+i\infty}\text{d}p\frac{f\left(s,p\right)}{p^2-3p+2}\operatorname{e}^{pu}\;,
\end{equation}
where $\operatorname{Re}\left(\beta\right)$ is larger than the fastest growing exponential term contained in $Y$. Eq. \eqref{eq:sol_laplace2} can be evaluated with the method of residues once the integrand is analytically continued in the left half complex plane. Since the poles are located in $p=1$ and $p=2$, the final result is
\begin{equation}
\label{eq:growth}
Y\left(s,u\right)= f\left(s,2\right)\operatorname{e}^{2u} - f\left(s,1\right)\operatorname{e}^u\;.
\end{equation}
One can easily check by direct substitution that this is indeed a solution of Eq. \eqref{eq:first1}-\eqref{eq:last1}.

In the limit $\left|\tau\right|\gg 1$ the outer radius of the flow is approximately $R\approx R_{\rm max}/\left|\tau\right|$. Using the definition of $f$ and the fact that $u\equiv\log\left|\tau/\tau_0\right|\approx \log\left(R_0/R\right)$, Eq. \eqref{eq:growth} can be written as
\begin{equation}
\label{eq:growth2}
Y= \left(\frac{R_0}{R}\right)^2\left[Z_0-Y_0\right] + \left(\frac{R_0}{R}\right) \left[2Y_0-Z_0\right] \;,
\end{equation}
which gives the complete dependence of the perturbation on $r/R$, through the functions $Y_0$ and $Z_0$, and on $z$.

When the flow is converging, $R$ is decreasing along $z$ and the fastest mode grows as $R^{-2}$. This mode is indeed growing a factor $R^{-1}$ faster than the typical amplitude of the unperturbed fields, which is proportional to $R^{-1}$. Hence, a strong instability develops in the vicinity of the recollimation nozzle.

Instead, when the flow is diverging $R$ is increasing along $z$ and the perturbed fields decay proportionally to $R^{-1}$. Since the ratio with the amplitude of the unperturbed fields remains constant, in this case the flow is stable. Finally, note that the growth of the unstable modes does not depend on the helicity $m$ of the perturbation.

\section{Conclusions}
\label{sec:conclusions}

We have performed a linear stability analysis of ultra-relativistic, highly magnetised outflows with a recollimation nozzle. We have considered perturbations propagating through the region close to the nozzle, where the flow is approximately conical. We have neglected the dynamical effect of the poloidal magnetic field, as appropriate in the causally disconnected regime. We have furthermore worked in the limit $kR\ll\gamma$, where $k$ is the wavenumber of the perturbation, $R$ is the outer radius and $\gamma$ is the Lorentz factor of the flow. This corresponds to the case where the wavelength of the perturbation is longer than $R$ in the proper frame of the flow. Our main results can be summarised as follows:

\begin{enumerate}
\item while $R$ shrinks from $R_1$ to $R_2<R_1$, the amplitude of non-axisymmetric perturbations propagating along the flow grows by a factor $R_1/R_2$ with respect to the unperturbed fields. This confirms the simple picture proposed by \citet{Lyubarsky2012}. Consider a purely toroidal magnetic field, so that the flow can be conceived as composed of coaxial magnetic coils. If one takes a small radial displacement $\delta R$ of two coils, the relative amplitude of the perturbed magnetic field is $\delta B/B \sim \delta R/R$. Assuming that $\delta R$ remains approximately constant as the outer radius $R$ shrinks, $\delta B/B$ indeed grows proportionally to $R^{-1}$.
\item the growth of the perturbations depends only on the geometry of the flow (namely $R_1$ and $R_2$), being instead independent of its Lorentz factor. Since the perturbations are amplified by an arbitrarily large factor in the limit $R_2\ll R_1$, the flow becomes violently unstable if the recollimation is strong enough. Instabilities excited while the flow is converging may therefore play an important role to dissipate the magnetic energy of the plasma, and understanding if/how much a realistic outflow is recollimated is crucial to assess their impact.
\end{enumerate}

In order to achieve a strong recollimation, the outflow needs first to get rid of the poloidal field that halts the compression through its magnetic pressure. The dynamical effect of the poloidal field is determined by its ratio with the toroidal field in the proper frame of the flow, namely $B_{\rm p}$ is negligible if $B_\phi/\gamma B_{\rm p}\gg 1$. Since $B_\phi/\gamma B_{\rm p}$ is proportional to $R^2$, a fast initial expansion allows the outer radius to shrink significantly thereafter. The amount of initial expansion is clearly connected to the properties of the central engine and to those of the confining medium. In general, we expect a significant expansion to be favoured when the initial (mostly magnetic) pressure of the ejected plasma largely exceeds that of the confining medium.

In the case of PWNe, where the pressure of the confining medium is extremely low, the instability studied in this paper is therefore a promising explanation for the observed gamma-ray flares. We refer to \citet{Lyubarsky2012} for a more extended discussion of this point in the context of the Crab nebula.

In the case of GRBs, the jet may expand significantly before being recollimated by the pressure of the stellar environment if it is launched by a magnetar-like engine (see for example the simulations of \citealt{BrombergTchekhovskoy2016}). Hence, as suggested by these authors, also in these objects the instabilities excited while the flow is converging may be responsible for the dissipation of the magnetic energy. This claim is supported by the fact that we find these instabilities not to be suppressed even in the ultra-relativistic regime.

Finally, note that we have not addressed the question of how the magnetic energy is actually converted into the observed radiation. Indeed, the dissipation of the magnetic energy is an intrinsically non-linear process whose study would require numerical simulations, which is out of the scope of the paper. Instead, we have simply assumed that the onset of MHD instabilities triggers the dissipation of the magnetic energy destroying the regular structure of the field lines. The actual dissipation may happen via reconnection and/or dissipation of MHD turbulence, as suggested by the results of numerical simulations (see for example \citealt{BrombergTchekhovskoy2016}).

\section*{Acknowledgements}

The authors acknowledge support from the Israeli Science Foundation under Grant No. 719/14.

\def\aap{A\&A}\def\aj{AJ}\def\apj{ApJ}\def\apjl{ApJ}\def\mnras{MNRAS}
\def\araa{ARA\&A}\def\physrep{PhR}\def\sovast{Sov. Astron.}\def\nar{NewAR}
\def\aapr{Astronomy \& Astrophysics Review}\def\apjs{ApJS}\def\nat{Nature}\def\na{New Astron.}
\def\prd{Phys. Rev. D}\def\pre{Phys. Rev. E}\def\pasp{PASP}
\bibliographystyle{mn2e}
\bibliography{2d}

\appendix
\section{Stability of the axisymmetric perturbations}
\label{sec:appendixA}

In the case $m=0$, in order to find an equation for $E_{1\phi}$ it is sufficient to consider (i) the $r$ and $z$ components of Eq. \eqref{eq:B1}; (ii) Eq. \eqref{eq:EB_lin}. These equations give respectively
\begin{align}
& B_{\rm 1r}+E_{1\phi}=\frac{i}{\omega}\frac{\partial E_{1\phi}}{\partial z} \label{eq:axisymmetric1}\\
& \frac{\partial}{\partial r}\left(rE_{1\phi}\right)=i\omega rB_{\rm 1z} \\
& B_{\rm 1r}+E_{1\phi}=-\frac{\tau r}{L}B_{\rm 1z}-\frac{1}{6}\left(1+3\tau^2\right)\frac{r^2}{L^2}E_{1\phi} \label{eq:axisymmetric3}\;.
\end{align}
Rearranging Eq. \eqref{eq:axisymmetric1}-\eqref{eq:axisymmetric3} it is simple to find a new equation containing $E_{1\phi}$ only:
\begin{equation}
\label{eq:axisymmetricEphi}
L\frac{\partial E_{1\phi}}{\partial z} = \tau r\frac{\partial E_{1\phi}}{\partial r} +\tau E_{1\phi} +i\frac{\omega r}{6}\left(1+3\tau^2\right)\frac{r}{L}E_{1\phi}\;.
\end{equation}
Neglecting the last therm, which is a factor $1/\gamma$ smaller than the others, and using the new variables $s\equiv -\log\left(r/R\right)$ and $\tau$, Eq. \eqref{eq:axisymmetricEphi} becomes
\begin{equation}
\left(1+\tau^2\right)\frac{\partial E_{1\phi}}{\partial \tau}=\tau E_{1\phi}\;,
\end{equation}
which is solved by
\begin{equation}
E_{1\phi}\left(s,\tau\right)=f\left(s\right)\sqrt{1+\tau^2}\;.
\end{equation}
Note that the first and the third boundary conditions of \eqref{eq:outer_boundary} are automatically satisfied if $f\left(0\right)=0$. In the limit $\left|\tau\right|\gg 1$ we have $E_{1\phi}\propto\left|\tau\right|\propto 1/R$, which means that the amplitude of these modes does not grow with respect to the unperturbed fields.

The missing equations describe the evolution of a second mode, which in the case $m=0$ becomes decoupled from the first one. The $\phi$ component of Eq. \eqref{eq:B1} gives
\begin{equation}
\label{eq:axisymmetric_Bphi}
B_{1\phi}-E_{\rm 1r}=-\frac{i}{\omega}\left[\frac{\partial E_{\rm 1r}}{\partial z}-\frac{\partial E_{\rm 1z}}{\partial r}\right]\;.
\end{equation}
The $r$ and $z$ components of Eq. \eqref{eq:j1} can be presented as
\begin{align}
\label{eq:axisymmetric_jr}
rj_{\rm 1r} & =-r\frac{\partial B_{1\phi}}{\partial z} -i\omega r\left(B_{1\phi}-E_{\rm 1r}\right) \\
\label{eq:axisymmetric_jz}
r\left(j_{\rm 1z}-\rho_1\right) & =\frac{\partial}{\partial r}\left[r\left(B_{1\phi}-E_{\rm 1r}\right)\right] -r\frac{\partial E_{\rm 1z}}{\partial z}\;,
\end{align}
where from the $z$ component we have subtracted Eq. \eqref{eq:rho1}. The $\phi$ component of Eq. \eqref{eq:j1} can be simply regarded as a definition of $j_{1\phi}$. Finally, Eq. \eqref{eq:rho1} is
\begin{equation}
\label{eq:axisymmetric_rho}
\frac{\partial}{\partial r}\left(rE_{\rm 1r}\right)=r\rho_1-i\omega rE_{\rm 1z} -r\frac{\partial E_{\rm 1z}}{\partial z}\;,
\end{equation}
while Eq. \eqref{eq:Ej_lin} and \eqref{eq:eulersq_lin} respectively give
\begin{align}
\label{eq:axisymmetric_Ej}
rj_{\rm 1r} & =-\frac{\tau r}{L}rj_{\rm 1z} +2\frac{\tau r}{L}E_{\rm 1r} -2E_{\rm 1z}\\
\label{eq:axisymmetric_euler}
r\left(j_{\rm 1z}-\rho_1\right) & =2\left(E_{\rm 1r}-B_{1\phi}\right) +\frac{1}{6}\left(1+3\tau^2\right)\frac{r^2}{L^2}\left(2E_{\rm 1r}-r\rho_1\right)\;.
\end{align}

Combining Eq. \eqref{eq:axisymmetric_jr}, \eqref{eq:axisymmetric_rho}, \eqref{eq:axisymmetric_Ej}, using Eq. \eqref{eq:axisymmetric_Bphi} to express $\partial E_{\rm 1r}/\partial z$ and neglecting the small corrections we find
\begin{equation}
\label{eq:axisymmetric_final1}
r\frac{\partial K}{\partial r} -2K = i\omega r\left[\frac{\tau^2 r^2}{L^2}E_{\rm 1r} -2\left(B_{1\phi}-E_{\rm 1r}\right)\right]\;,
\end{equation}
which shows that
\begin{equation}
K\equiv E_{\rm 1z}-\frac{\tau r}{L}E_{\rm 1r}
\end{equation}
is of the order of $1/\gamma^2$. A second equation is obtained combining Eq. \eqref{eq:axisymmetric_jz} and \eqref{eq:axisymmetric_euler}, using Eq. \eqref{eq:axisymmetric_jr} and \eqref{eq:axisymmetric_Ej} to express $2E_{\rm 1r}-\rho_1$, and neglecting terms of order higher than $1/\gamma^2$. In the limit $\left|\tau\right|\gg 1$ we get
\begin{equation}
\label{eq:axisymmetric_final2}
\frac{\partial}{\partial r}\left[r\left(B_{1\phi}-E_{\rm 1r}\right)\right] +2\left(B_{1\phi}-E_{\rm 1r}\right) = \frac{\tau^2 r^2}{L^2}E_{\rm 1r} +\frac{1}{2}\frac{\tau r}{L} r\frac{\partial E_{\rm 1r}}{\partial z}\;.
\end{equation}
Finally, the last equation is simply obtained imposing that the terms of the order of $1/\gamma$ on the right hand side of Eq. \eqref{eq:axisymmetric_Bphi} vanish, which gives
\begin{equation}
\label{eq:axisymmetric_final3}
\frac{\partial E_{\rm 1r}}{\partial z}=\frac{\partial}{\partial r}\left(\frac{\tau r}{L}E_{\rm 1r}\right)\;.
\end{equation}
We now need to solve the three equations \eqref{eq:axisymmetric_final1}, \eqref{eq:axisymmetric_final2}, \eqref{eq:axisymmetric_final3} for the three unknown functions $E_{\rm 1r}$, $K$, $B_{1\phi}-E_{\rm 1r}$.

Introducing the variables $s\equiv-\log\left(r/R\right)$ and $u\equiv\log\left|\tau/\tau_0\right|$, Eq. \eqref{eq:axisymmetric_final3} becomes
\begin{equation}
\frac{\partial E_{\rm 1r}}{\partial u}=E_{\rm 1r}\;,
\end{equation}
which is solved by
\begin{equation}
E_{\rm 1r}=g\left(s\right)\operatorname{e}^u\;.
\end{equation}
It is now simple to realise that the solution of Eq. \eqref{eq:axisymmetric_final1} and \eqref{eq:axisymmetric_final2} is of the form
\begin{equation}
K=i\omega r\frac{\tau^2 r^2}{L^2} h\left(s\right)\operatorname{e}^u \qquad B_{1\phi}-E_{\rm 1r}=\frac{\tau^2 r^2}{L^2} l\left(s\right)\operatorname{e}^u\;,
\end{equation}
where the functions $g$ and $l$ can be calculated from $h$ using respectively (i) $2g = h'' -6h' +5h$ and (ii) $4l = h'' -4h' +3h$. Note that the second boundary condition of \eqref{eq:outer_boundary} is automatically satisfied if $h\left(0\right)=0$. Since in the limit $\left|\tau\right|\gg 1$ the amplitude of all the perturbations is proportional to $\left|\tau/\tau_0\right|\approx R_0/R$, we can finally conclude that the axisymmetric modes do not grow with respect to the unperturbed fields.

\end{document}